\documentclass[letterpaper,english,reprint,aps,prl,superscriptaddress,showpacs,longbibliography]{revtex4-2}
\usepackage[utf8]{inputenc}
\usepackage{amsmath}
\usepackage{amssymb}
\usepackage{graphicx}
\usepackage{soul}
\usepackage{color, xcolor}
\makeatletter

\pdfpageheight\paperheight
\pdfpagewidth\paperwidth

\usepackage[colorlinks,linkcolor=blue,anchorcolor=black,citecolor=blue,urlcolor=blue]{hyperref}

\makeatother

\usepackage{babel}

\begin{document}
	\title{Highly sensitive measurement of a megahertz rf electric field with a Rydberg-atom sensor}    
	\author{Bang Liu}
	\affiliation{Key Laboratory of Quantum Information, University of Science and Technology
		of China, Hefei, Anhui 230026, China.}
	\affiliation{Synergetic Innovation Center of Quantum Information and Quantum Physics,
		University of Science and Technology of China, Hefei, Anhui 230026,
		China.}
	\author{Li-Hua Zhang}
	\affiliation{Key Laboratory of Quantum Information, University of Science and Technology
		of China, Hefei, Anhui 230026, China.}
	\affiliation{Synergetic Innovation Center of Quantum Information and Quantum Physics,
		University of Science and Technology of China, Hefei, Anhui 230026,
		China.}
	\author{Zong-Kai Liu}
	\affiliation{Key Laboratory of Quantum Information, University of Science and Technology
		of China, Hefei, Anhui 230026, China.}
	\affiliation{Synergetic Innovation Center of Quantum Information and Quantum Physics,
		University of Science and Technology of China, Hefei, Anhui 230026,
		China.}
	\author{Zheng-Yuan Zhang}
	\affiliation{Key Laboratory of Quantum Information, University of Science and Technology
		of China, Hefei, Anhui 230026, China.}
	\affiliation{Synergetic Innovation Center of Quantum Information and Quantum Physics,
		University of Science and Technology of China, Hefei, Anhui 230026,
		China.}
	\author{Zhi-Han Zhu}
	\affiliation{Wang Da-Heng Collaborative Innovation Center for Science of Quantum
		Manipulation and Control, Heilongjiang Province and Harbin University
		of Science and Technology, Harbin 150080, China}
	\author{Wei Gao}
	\affiliation{Wang Da-Heng Collaborative Innovation Center for Science of Quantum
		Manipulation and Control, Heilongjiang Province and Harbin University
		of Science and Technology, Harbin 150080, China}
	\author{Guang-Can Guo}
	\affiliation{Key Laboratory of Quantum Information, University of Science and Technology
		of China, Hefei, Anhui 230026, China.}
	\affiliation{Synergetic Innovation Center of Quantum Information and Quantum Physics,
		University of Science and Technology of China, Hefei, Anhui 230026,
		China.}
	\author{Dong-Sheng Ding}
	\email{dds@ustc.edu.cn}
	
	\affiliation{Key Laboratory of Quantum Information, University of Science and Technology
		of China, Hefei, Anhui 230026, China.}
	\affiliation{Synergetic Innovation Center of Quantum Information and Quantum Physics,
		University of Science and Technology of China, Hefei, Anhui 230026,
		China.}
	\affiliation{Wang Da-Heng Collaborative Innovation Center for Science of Quantum
		Manipulation and Control, Heilongjiang Province and Harbin University
		of Science and Technology, Harbin 150080, China}
	\author{Bao-Sen Shi}
	\email{drshi@ustc.edu.cn}
	
	\affiliation{Key Laboratory of Quantum Information, University of Science and Technology
		of China, Hefei, Anhui 230026, China.}
	\affiliation{Synergetic Innovation Center of Quantum Information and Quantum Physics,
		University of Science and Technology of China, Hefei, Anhui 230026,
		China.}
	\date{\today}

	\begin{abstract}
		Rydberg atoms have great potential in electric field measurement and
		have an advantage with a large frequency bandwidth from the kHz to
		the THz scale. However, the sensitivity for measuring a weak MHz electric
		field signal is limited by the spectroscopic resolution, because the
		weak electric field induces only a small perturbation of the population
		and energy level shift of the Rydberg atoms. Here, we report highly
		sensitive measurement of a weak MHz electric field using electromagnetically
		induced transparency with Rydberg atoms in a thermal atomic system.
		Using the heterodyne method on a 30-MHz electric field, we successfully
		measure the minimum electric field strength to be \textcolor{black}{37.3 $\mathrm{\mu V/cm}$}
		with a sensitivity up to $-65$ dBm/Hz and a linear dynamic range
		over 65 dB. Additionally, we measure an amplitude-modulated signal
		and demodulate the signal with a fidelity over 98\%. This work extends
		the sensitivity of atomic sensors for measuring MHz electric fields,
		which advances atomic electric field-sensing technology. 
	\end{abstract}

	\maketitle
	
	\section{INTRODUCTION}
	
	Rydberg atoms have been widely used for electric field measurement
	owing to their large electric dipole moments and polarizabilities
	\citep{gallagher2006rydberg}. Their various energy levels, covering
	ranges from the kHz to the THz scale, make it easier to expand their
	operating bands than those of conventional antennas without changing
	the sensing device, so Rydberg atoms shows great potential for measuring
	electric fields. Initially, Rydberg atoms were detected with ionization
	pulses, but those were destructive \citep{osterwalder1999using}.
	Then, the use of electromagnetically induced transparency (EIT) with
	Rydberg atoms \citep{mohapatra2007coherent,kubler2010coherent} was
	proposed as a direct, non-destructive method to probe Rydberg energy
	levels. Using the Rydberg-EIT method, researchers have studied the
	spectra of Rydberg atoms modulated by radio frequency (RF) electric
	fields and obtained the strengths of the applied RF electric fields
	\citep{sedlacek2012microwave,jing2020atomic}. Electric field measurement
	techniques based on Rydberg atoms have progressed from weak to strong
	fields and successfully measured field strengths from the nV/m \citep{jing2020atomic,gordon2019weak,kumar2017rydberg}
	to the kV/m scale \citep{anderson2016optical,paradis2019atomic}.
	Furthermore, an electric field with a large frequency bandwidth can
	be measured from below 1 kHz \citep{jau2020vapor} to tens of GHz
	\citep{downes2020full,gordon2014millimeter,wade2017real}. All these
	efforts are aimed at establishing an electric field measurement with
	an atomic standard.
	
	Researchers are also applying Rydberg atomic measurements of electric
	fields to such areas as RF polarization \citep{sedlacek2013atom},
	subwavelength imaging \citep{holloway2014sub,holloway2017atom}, and
	digital communications \citep{meyer2018digital,song2019rydberg,liu2022deep}.
	This is especially significant for technologies involving $\sim$MHz
	electric fields such as shortwave international and regional broadcasting
	and aviation air-to-ground communications, because of the long wavelengths
	and long propagation distances of these fields. However, the Chu limit
	of conventional antennas \citep{chu1948physical} restricts the channel
	capacity when the antenna is much smaller than the wavelength of the
	electromagnetic wave. In contrast, a sensor based on Rydberg atoms is
	not limited by size, the vapor cell used is generally a few cm long,
	and the data capacity of the sensor far exceeds that of a conventional
	antenna of the same size \citep{cox2018quantum}. Therefore, it makes
	sense to develop a technique that measures a MHz electric field with
	high sensitivity using Rydberg atoms.
	
	Although electric fields can be measured over a wide range of frequencies
	based on Rydberg atoms, it is rare to measure a weak MHz signal. It
	is difficult to measure the shift of an EIT spectrum because the weak
	field causes very small perturbations to the energy level, making
	the EIT spectrum change very little. Some groups have studied related
	problems \citep{jiao2016spectroscopy,miller2016radio,anderson2016optical},
	but they were more focused on strong fields. In a strong field, the
	Rydberg atomic energy levels have a large Stark shift, which leads
	to mixing of many different energy levels. Floquet theory is used
	to calculate the Rydberg-EIT spectrum of a strong field. Information
	on the applied electric field can be extracted by comparing the calculated
	Rydberg-EIT spectrum of the field with the experimentally measured
	spectrum. Researchers have used this method to measure strong fields
	with amplitudes greater than 5 kV/m \citep{paradis2019atomic}, but
	it is difficult to measure a weak electric field with the method because
	the minimum measurable field strength is only 0.1 V/m.
	
	\begin{figure*}[t]
		\includegraphics[width=1.8\columnwidth]{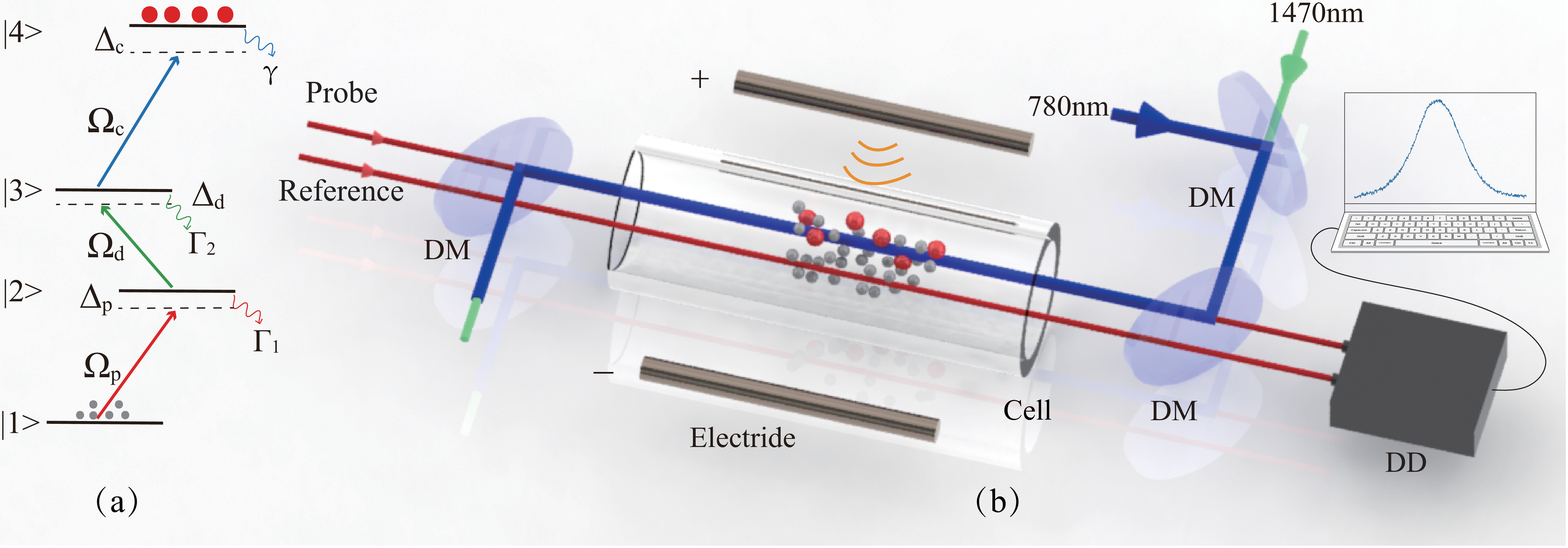}
		
		\caption{(a) Ladder-type four-level atomic energy diagram consisting of a ground
			state $\left|1\right\rangle $, two low-lying excited states $\left|2\right\rangle $
			and $\left|3\right\rangle $, and a Rydberg state $\left|4\right\rangle $.
			An 852-nm probe light drives the transition $\left|1\right\rangle =\left|6S_{1/2},F=4\right\rangle \rightarrow\left|2\right\rangle =\left|6P_{3/2},F=5\right\rangle $,
			a 1470-nm dressing light couples the transition $\left|2\right\rangle =\left|6P_{3/2},F=5\right\rangle \rightarrow\left|3\right\rangle =\left|7S_{1/2},F=4\right\rangle $,
			and a 780-nm coupling light drives the transition $\left|3\right\rangle =\left|7S_{1/2},F=4\right\rangle \rightarrow\left|4\right\rangle =\left|55P_{3/2}\right\rangle $
			of cesium atoms. (b) Overview of the experimental setup. The probe
			light and the reference light propagate in parallel through a Cs vapor
			cell. The probe light (red) overlaps the counter-propagating coupling
			light (blue) and dressing light (green) to form an EIT configuration.
			The transmission difference between the probe and reference lights
			is detected by a differencing photodetector. Two electrode rods are
			placed parallel to each other on both sides of the vapor cell 4 cm
			apart. Labels: DM - dichroic mirror; DD - differencing photodetector.}
		
		\label{Figure1} 
	\end{figure*}
	
	In this study, \textcolor{black}{we focus on the highly sensitive measurement of a MHz RF electric field using a hot vapor cell. The outside electric field can be coupled directly to the atoms without the need for additional devices. As the MHz electric field does not couple any transitions between two Rydberg levels in our experiment, we use the ac Stark shift to measure the MHz electric field.} And we use the heterodyne technique \citep{jing2020atomic,meyer2021waveguide}
	to amplify the system response to a weak signal electric field by
	applying a local electric field. We achieve measurement of a 30-MHz
	electric field with a sensitivity of $-65$ dBm/Hz and dynamic range
	of 65 dB. In addition, we demonstrate measurement of an amplitude-modulated
	(AM) 1-kHz signal electric field based on Rydberg atoms with a fidelity
	over 98\%. Our work helps applications of Rydberg atomic sensors such
	as long-distance communication, over-the-horizon radar, and radio
	frequency identification (RFID). 
	
	\section{EXPERIMENT setup}
	
	Figure \ref{Figure1}(a) is the energy level diagram of a cesium (Cs)
	atom, where $\left|1\right\rangle $ is the ground state, $\left|2\right\rangle $
	and $\left|3\right\rangle $ are two low-lying excited states, and  $\left|4\right\rangle $
	is the Rydberg state of the Cs atom. The experimental setup is depicted
	in Fig. \ref{Figure1}(b). The probe light passes through a 7-cm vapor
	cell in parallel with the reference light, and the dressing light
	and the coupling light propagate backwards from the probe light. The
	probe light is focused into a cell ($1/e^{2}$-waist radius of approximately
	200 $\mathrm{\mu m}$ ) and couples the ground state $\left|1\right\rangle =\left|6S_{1/2},F=4\right\rangle $
	to the intermediate state $\left|2\right\rangle =\left|6P_{3/2},F=5\right\rangle $
	with Rabi frequency $\Omega_{p}$. The dressing light is focused into
	the cell ($1/e^{2}$-waist radius of approximately 500 $\mathrm{\mathrm{\mu m}}$)
	and couples the two intermediate states $\left|2\right\rangle =\left|6P_{3/2},F=5\right\rangle $
	and $\left|3\right\rangle =\left|7S_{1/2},F=4\right\rangle $ with
	Rabi frequency $\Omega_{d}$. The coupling light ($1/e^{2}$-waist
	radius of approximately 500 $\mathrm{\mu m}$) drives the transition
	from $\left|3\right\rangle =\left|7S_{1/2},F=4\right\rangle $ to
	the Rydberg state $\left|4\right\rangle =\left|55P_{3/2}\right\rangle $
	with detuning $\Delta_{c}$ and Rabi frequency $\Omega_{c}\sim2\pi\times4$
	MHz. A four-energy-level structure is used here to avoid the use of
	lasers below 580 nm, thus avoiding the ionization shielding effect
	of Cs atoms in the atomic vapor cell due to photoelectric ionization
	\citep{bason2010enhanced,viteau2011rydberg,xu1996photoatomic}. \textcolor{black}{In the three-photon EIT scheme, lasers are simple commercial lasers which can be acquired easily and we do not require complex Frequency-doubling lasers, which are very expensive.} We
	use two RF signal sources to generate the RF wave, one as the local
	oscillator (LO) and the other as the signal, where the LO has a small
	detuning from the signal. The signal and LO electric field pass through
	a transmission line through an RF power splitter to the electrode
	rods. The transmission of the probe and reference lights is detected
	by a balanced photodetector.
	
	\section{RESULTS AND DISCUSSION}
	
\begin{figure*}[t]
		\includegraphics[width=2\columnwidth]{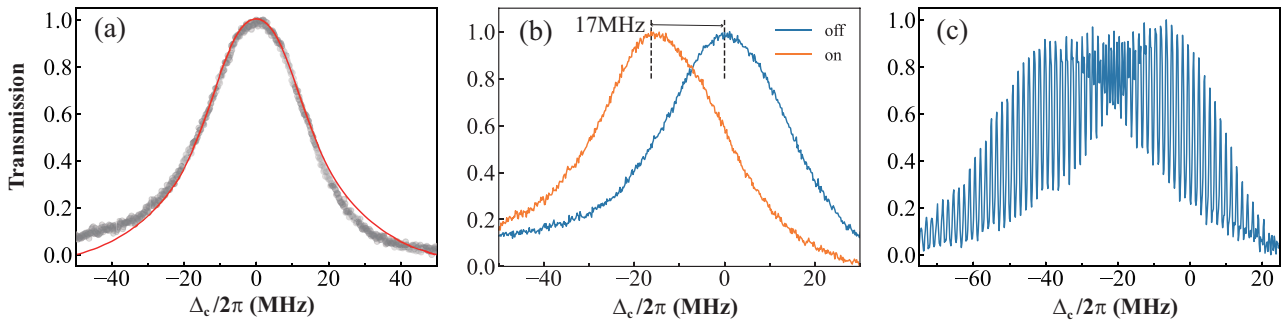}
		
		\caption{(a) Three-photon EIT spectrum of Cs atoms. The gray dots are the experimental
			data, and the red line is the theoretical result. (b) EIT spectrum
			in an external RF electric field. The RF electric field causes the
			EIT spectrum to shift to the left. \textcolor{black}{The blue is the spectrum when
			the RF electric field is turned off, and the orange correspond to
			the spectrum when the RF electric field is turned on and P= 8 dBm}. (c) EIT spectrum of the Cs atoms when two electric
			fields are applied, where $f_{\mathrm{sig}}=30$ MHz and $f_{\mathrm{LO}}=30.05$
			MHz. The EIT spectrum is modulated with $\Delta\omega=50$ kHz by
			the beat signals of the two electric fields. \textcolor{black}{The transmission in the pictures is normalized.}}
		
		\label{Figure2} 
	\end{figure*}

	First, we experimentally studied the three-photon EIT in the Cs vapor
	cell involving the Rydberg state \citep{thaicharoen2019electromagnetically}.
	We fixed the frequencies of the probe light and the dressing light
	such that each resonated with the corresponding transitions. Figure
	\ref{Figure2}(a) shows the EIT spectrum of the Cs atoms obtained
	by scanning the detuning $\Delta_{c}$ of the coupling light. 
	
	When $\Delta_{c}=0$, the spectrum exhibits a narrow transmission
	window, which is a typical EIT feature. To model our system, we consider
	the four-level system in Fig. \ref{Figure1}(a). The detunings of
	the probe light, dressing light, and coupling light are denoted as
	$\Delta_{p}$, $\Delta_{d}$, and $\Delta_{c}$, respectively; and
	$\Gamma_{1}$, $\Gamma_{2}$, and $\gamma$ are the corresponding
	decay rates of the atomic states $\left|2\right\rangle $, $\left|3\right\rangle $,
	and $\left|4\right\rangle $, respectively. Using the rotating wave
	approximation, the Hamiltonian of the system in the interaction picture
	can be written as 
	\begin{alignat}{1}
		H_{I} & =\hbar\Delta_{p}\sigma_{22}+\hbar\left(\Delta_{p}+\Delta_{d}\right)\sigma_{33}+\hbar\left(\Delta_{p}+\Delta_{d}+\Delta_{c}\right)\sigma_{44}\nonumber \\
		& -\hbar/2\left(\Omega_{p}\sigma_{12}+\Omega_{d}\sigma_{23}+\Omega_{c}\sigma_{34}+\mathrm{H.C.}\right)\label{eq:1}
	\end{alignat}
	where $\sigma_{ij}=\left|i\right\rangle \left\langle j\right|$ $\left(i,j=1,2,3,4\right)$
	are the atomic transition operators. Considering spontaneous radiation,
	the system satisfies the Lindblad equation
	
	\begin{eqnarray}
		\dot{\rho} & = & -\frac{i}{\hbar}\left[H,\rho\right]+\mathcal{L}\left(\Gamma_{21}\right)+\mathcal{L}\left(\Gamma_{32}\right)+\mathcal{L}\left(\Gamma_{43}\right)\label{eq:2}
	\end{eqnarray}
	where $\rho$ is the density matrix of system, and $\mathcal{L}\left(\Gamma_{ij}\right)=\Gamma_{ij}/2\left(2\sigma_{ji}\rho\sigma_{ij}-\sigma_{ii}\rho-\rho\sigma_{ii}\right)$
	is the Lindblad operator with $\Gamma_{21}=\Gamma_{1}$, $\Gamma_{32}=\Gamma_{2}$,
	and \textcolor{black}{$\Gamma_{43}=\gamma$}. Considering the Doppler effect due to the
	thermal motion of the atoms, we correct the detunings to $\Delta_{p}=\Delta_{p}-k_{p}v$, $\Delta_{d}=\Delta_{d}+k_{d}v$, and $\Delta_{c}=\Delta_{c}+k_{c}v$,
	where $k_{p}$, $k_{d}$, and $k_{c}$ are the corresponding wave
	vectors, and $v$ denotes the atom velocity. The steady-state solution
	of the system density matrix can be obtained by solving Eq. \ref{eq:2}.
	The susceptibility 
	\begin{equation}
		\chi_{21}=-\int\frac{2N\left(v\right)|d_{21}|^{2}}{\hbar\epsilon_{0}\Omega_{p}}\rho_{21}\left(v\right)dv
	\end{equation}
	is further obtained, where $N\left(v\right)=\frac{N_{0}}{u\sqrt{\pi}}\exp\left(-v^{2}/u^{2}\right)$,
	$u=\left(2k_{B}T/M\right)^{-1/2}$ is the most probable speed, $N_{0}$
	is the atomic density, $d_{21}$ is the transition dipole matrix element
	between $\left|2\right\rangle $ and $\left|1\right\rangle $, $k_{B}$
	is the Boltzmann constant, $T$ is the temperature of the cell, and
	$M$ is the mass of a Cs atom. The absorption coefficient can be calculated
	using the imaginary part of the susceptibility, $\alpha=k_{p}\mathrm{Im}\left(\chi_{21}\right)$.
	The transmission is calculated as exp$\left(-\alpha l\right)$ and
	plotted as a function of coupling detuning $\Delta_{c}$ in Fig. \ref{Figure2}(a),
	where $l$ is the length of the cell. The probe light transmission
	obtained by solving the master equation (red) agrees well with the
	experiment (\textcolor{black}{gray}). 
	
	We then investigated the effect of an RF electric field on the atomic
	EIT spectrum. In the experiment, only one signal source was turned
	on, and its frequency was set to 30 MHz with an output power of \textcolor{black}{8
	dBm}. The RF electric field signal was applied to the electrode rods
	along the transmission line. Turning on the RF electric field caused
	the entire EIT spectrum to shift to the left, and \textcolor{black}{the spectrum is shifted by about 17 MHz}, as shown
	in Fig. \ref{Figure2}(b). This can be explained by the ac Stark shift
	\citep{delone1999ac,gallagher2006rydberg} $\delta=-\frac{1}{2}\alpha E^{2}$,
	where $\alpha$ is the polarizability and $E$ the amplitude of the
	electric field. The Rydberg state is more sensitive to the external
	electric field owing to the higher polarizability, which induces a
	large energy shift. The ground and intermediate states produce a small
	energy shift because of the lower polarizability, so we can ignore
	it. When considering the perturbation of the Rydberg state by the
	external electric field, the detuning of coupling light in Eq. \ref{eq:1}
	is updated to $\Delta_{c}=\Delta_{c}+\delta$. The EIT spectrum obviously
	shifts to the left as $\delta<0$. \textcolor{black}{Using the ac Stark effect, we can calibrate the magnitude of the electric field inside the vapor cell. With the Alkali Rydberg Calculator (ARC) package \citep{SIBALIC2017319}, we can calculate the polarizability of the Rydberg atoms to be 2500.68 MHz $\mathrm{ cm^{2}/V^{2}}$, and we can obtain the electric field as 167 mV/cm at a stark shift of 17 MHz. The electric field felt by the atoms is significantly smaller compared to the measured voltage between the electrode rods, due to the shielding effect of the vapor cell.} Note that the condition we consider
	is different from the resonant situation in \citep{sedlacek2012microwave},
	which leads to the Autler-Townes (AT) \textcolor{black}{splitting} \citep{autler1955stark}.
	Because the applied RF electric field is far from resonance, it does
	not couple the Rydberg state transition. The electric field just causes
	an ac Stark shift of the Rydberg energy level, which leads to a shift
	of the EIT resonance peak.
	
	Next, we demonstrated measurement of a weak MHz RF electric field
	based on Rydberg atoms through applying an LO electric field using
	a heterodyne method, which is our main work. As mentioned above, the
	MHz RF electric field can cause an energy shift of the Rydberg state,
	and further shift the EIT spectrum of the cesium atoms. When two electric
	fields with different frequencies are applied, each perturbs the Rydberg
	energy level differently, so the EIT spectrum takes on a new character.
	We turned on two signal sources and set their frequencies to 30 MHz
	and 30.05 MHz, respectively; the EIT spectrum is shown in Fig. \ref{Figure2}(c).
	First, the EIT spectrum is clearly modulated, and the measured modulation
	frequency is exactly equal to the difference between the LO and signal
	electric field. In addition, the EIT spectrum is shifted to the left.
	The total electric field is $\mathbf{E}_{\mathrm{LO}}+\mathbf{E}_{\mathrm{sig}}$,
	and the Stark shift caused by the electric field is 
	\begin{equation}
		\delta=-\frac{1}{2}\alpha\left(\mathbf{E}_{\mathrm{LO}}+\mathbf{E}_{\mathrm{sig}}\right)^{2}\label{eq:3}
	\end{equation}
	Here we set the LO to be phase-synchronized with the signal in our
	experiment, so we let $\phi_{\mathrm{LO}}=\phi_{\mathrm{sig}}$, where
	$\Delta\omega$ represents the detuning of the LO and signal. When
	Eq. \ref{eq:3} is expanded, there are some fast-varying terms. By
	averaging over time, we get 
	\begin{align}
		\bar{\delta} & =\bar{\delta}_{0}-\frac{1}{2}\alpha\left[E_{\mathrm{LO}}E_{\mathrm{sig}}\cos(\Delta\omega*t)\right]\label{eq:4}
	\end{align}
	where $\bar{\delta}_{0}=-\frac{1}{4}\alpha\left(E_{\mathrm{LO}}^{2}+E_{\mathrm{sig}}^{2}\right)$
	is the average ac Stark shift caused by the LO and signal electric
	field.
	
	\begin{figure}[t]
		\includegraphics[width=0.9\columnwidth]{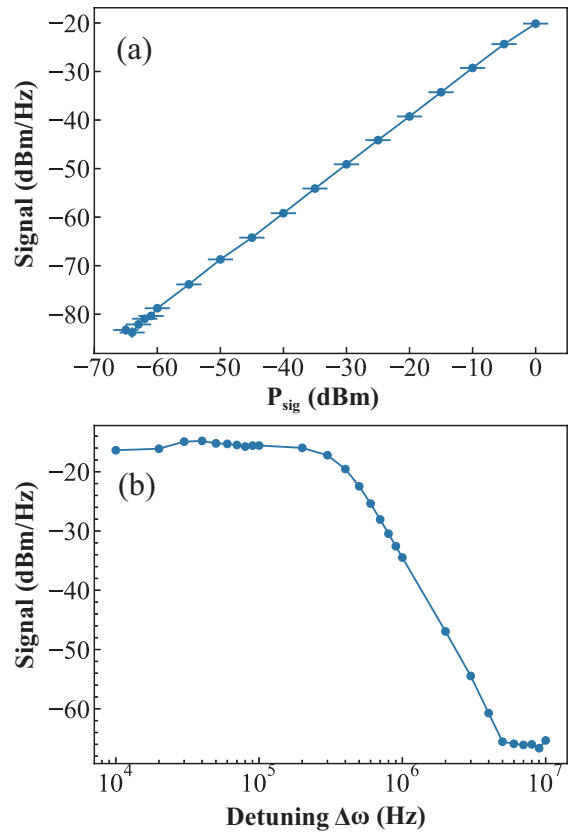}
		
		\caption{(a) Dynamic range of the system with $P_{\mathrm{LO}}=7$ dBm and
			$\Delta\omega=50$ kHz. The beat signal strength shows a linear relationship
			with input power in a range of 65 dB. (b) Instantaneous bandwidth
			of the system with $P_{\mathrm{LO}}=7$ dBm and $P_{\mathrm{sig}}=0$
			dBm. As the detuning of the signal electric field and the local oscillator
			electric field increases, the beat signal strength keeps decreasing.
			Considering the negative detuning, we get a 3-dB bandwidth of 0.8
			MHz.}
		
		\label{Figure3} 
	\end{figure}
	
	The energy shift is modulated by the beat frequency in Eq. \ref{eq:4},
	so we can see the beat frequency signal in the EIT spectrum. Additionally,
	information on the magnitude of the signal electric field can be extracted
	by measuring the beat frequency components of the probe light. In
	the experiment, the output of the two signal sources was applied to
	the electrode rods after the RF power splitter. The LO frequency was
	set to 30.05 MHz and the signal frequency to 30 MHz, and the LO power
	was $P_{\mathrm{LO}}=7$ dBm. The coupling light was fixed at an optimal
	operating point, and the change in probe light intensity with time
	was monitored through the digital oscilloscope. Then, we used the
	spectrum analyzer to measure the intensity of the beat signal. The
	beat signal intensity is shown in Fig. \ref{Figure3}(a) for different
	signal electric field strengths. The intensity of the received beat
	signal is approximately proportional to the strength of the applied
	signal electric field, as described by Eq. \ref{eq:4}. We achieved
	a dynamic range of 65 dB for measuring a 30-MHz RF electric field,
	and the power sensitivity was up to $-65$ dBm/Hz. \textcolor{black}{Using the previous calibration relationship between the electric field in the electrode rods and the input power, which we obtained in Fig. \ref{Figure2}(b)},
	we calculate the sensitivity for the electric field strength to be
	\textcolor{black}{37.3 $\mathrm{\mu V/cm/Hz^{1/2}}$}. The instantaneous bandwidth of the
	system is shown in Fig. \ref{Figure3}(b). The instantaneous bandwidth
	of the system \textcolor{black}{$B$} reaches 0.8 MHz considering the negative detuning of
	the signal with the LO. \textcolor{black}{The bandwidth of system mainly depends on the probe photon scattering rate of the intermediate atomic resonance (of order 10 MHz) \citep{Meyer_2020}. Here we  define a parameter $\eta=B / \omega_{c}$, i.e. the ratio of bandwidth and carrier frequency, to measure the relative bandwidth. Compared with \citep{jing2020atomic}, the parameter in our case is much larger than theirs. As they use GHz electric field to couple the Rydberg levels resonantly. Once the frequency of signal is deviates too much from the frequency of the LO, the response of the Rydberg atoms to signal will be weaker. However, here we use the ac Stark shift which is a non-resonant effect, and atoms  can respond to the electric field with a lager bandwidth.}
	
	\begin{figure*}[t]
		\includegraphics[width=2\columnwidth]{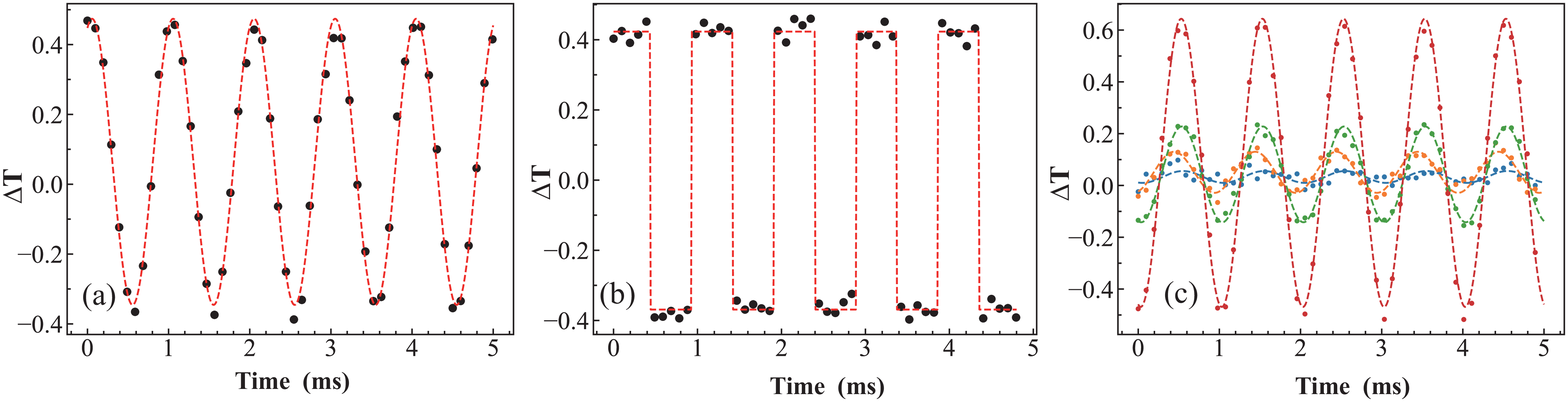}
		
		\caption{Results of demodulation of a 30-MHz carrier wave. (a) and (b) are
			the signals obtained by demodulating the carrier with sinusoidal modulation
			and square wave modulation, respectively, with a modulation frequency
			of 1 kHz. The signal is well restored with 98\% fidelity for both
			sinusoidal and square wave modulations. The black dots represent the
			signal after demodulation using Rydberg atoms, and the red dashed
			line is the modulated signal. \textcolor{black}{(c) Signals obtained by Rydberg atomic demodulation at different modulation depths, 2\% (blue), 5\% (green), 10\% (orange), 30\% (red).}}
		
		\label{Figure4} 
	\end{figure*}
	
	We have demonstrated short-wave communication using Rydberg atoms
	based on measurement of a MHz electric field using Rydberg atoms.
	RF electric fields can be used to carry information through sound
	signals and transmit information over great distances using the propagation
	properties of the carrier wave. Experimentally, we used a signal source
	to generate a 30-MHz RF electric field and amplitude-modulate it with
	a modulation frequency of 1 kHz \citep{anderson2020atomic,holloway2019real,jiao2019atom}. \textcolor{black}{The modulation depth is 30$\%$ and the carrier power is 7 dBm}. The transmission of the probe light
	was recorded by the digital oscilloscope when the coupling light was
	locked at the optimal operating point. As shown in Fig. \ref{Figure4},
	the modulation information is better extracted for both sinusoidal
	and square wave modulations. \textcolor{black}{We use $A \sin (\omega t+\varphi)+B$ to perform the fit where $\omega=2 \pi \times 1$ kHz and then calculate the deviation to obtain the fidelity, mainly to measure the degree of signal distortion after atomic demodulation. Compared to modulated signal, the fidelity of the demodulated signal
	reaches 98\%. Experimentally at different modulation depths, we extracted modulated signals from the spectra as shown in the Fig. \ref{Figure4}(c). Obviously, we can see that as the modulation depth increases, the clearer the extracted signal is.} The Rydberg atom acts as a demodulator without the need
	for complex electronic demodulation devices. The modulation signal
	is carried directly in the transmission of the probe light, and we
	only need to record it using the photoelectric detector. The carrier
	band we are concerned with is the MHz range. Although its signal bandwidth
	has a limit, this is not a major issue in areas such as broadcasting. \textcolor{black}{Actually, broadcasting is generally for long distance transmission, and most of the information transmitted is sound signals, which are mostly in kHz, and the instantaneous bandwidth we obtained in the experiment is 0.8 MHz, which is sufficient for transmitting sound information.}
	We are more concerned about the effect of the sensor size. When entering
	the electrically small regime \citep{cox2018quantum}, the efficiency
	of conventional antennas is greatly reduced while Rydberg atom-based
	sensors are not affected.
	
	\section{CONCLUSION}
	
	In conclusion, we investigated the perturbation of the Rydberg energy
	level by a MHz electric field, and realized measurement of a weak
	electric field signal using the heterodyne method. We achieved measurement
	of a 30-MHz electric field with a 65-dB dynamic range and a sensitivity
	reaching $-65$ dBm/Hz. On this basis, we demonstrated the recovery
	of an AM signal with 98\% fidelity. We have improved the performance
	of the Rydberg atom sensor in measuring electric fields in the MHz
	band, which will facilitate implementation of a wide-band Rydberg
	atom receiver. Measurements of electric fields at lower frequencies
	are also being explored \citep{jau2020vapor}, although many difficulties
	are encountered. It will be beneficial to use Rydberg atoms in more
	situations such as long-wave radar and submarine communication. 
	
	\section*{acknowledgments}
	
	\begin{acknowledgments}
		We acknowledge funding from the National Key R\&D Program of China
		(Grant No. 2017YFA0304800), the National Natural Science Foundation
		of China (Grant Nos. U20A20218, 61525504, and 61435011), the Anhui
		Initiative in Quantum Information Technologies (Grant No. AHY020200), the major science and technology projects in Anhui Province (Grant No. 202203a13010001)
		and the Youth Innovation Promotion Association of the Chinese Academy
		of Sciences (Grant No. 2018490). 
	\end{acknowledgments}

\end{document}